\documentclass[aps,pra,twocolumn,longbibliography] {revtex4-2}
\usepackage{amsmath,amssymb,graphicx,amsfonts}
\usepackage{multirow}
\usepackage[export]{adjustbox}
\usepackage{mhchem}
\usepackage{times}
\usepackage[colorlinks=true,linkcolor=blue,urlcolor=blue,citecolor=blue,breaklinks=true]{hyperref}
\usepackage{comment}
\usepackage{graphicx}
\usepackage[caption=false]{subfig}  
\usepackage{graphicx}

\def\da{\dagger}

\def\eq#1{Eq.~(\ref{#1})}
\def\da{\dagger}

\date\today
\begin{document}

\title{Driven dynamics of an attractive Bose polaron}

\author{Saptarshi Majumdar}
\affiliation{Univ Toulouse, CNRS, Laboratoire de Physique Th\'{e}orique, Toulouse, France}

\author{Aleksandra Petkovi\'{c} }
\affiliation{Univ Toulouse, CNRS, Laboratoire de Physique Th\'{e}orique, Toulouse, France}

\begin{abstract}
We study the out-of-equilibrium dynamics of an impurity driven by a constant external force through a system of {homogeneous} weakly-interacting bosons in one spatial dimension. The impurity-boson interaction is assumed to be attractive. We show that the impurity exhibits drifted Bloch oscillations in a wide range of forces in the absence of a lattice. We characterize the dynamical response of the host bosons and explain the mechanism underlying the Bloch oscillations. We analyse the behavior of the drift velocity, the Bloch amplitude and the time period of oscillations in a wide range of forces and other system parameters. In contrast to the case of repulsive impurity-boson interaction, the drift velocity exhibits a sub-linear dependence on a weak applied force, $V_d\sim {F}^{\alpha}$ with a positive exponent $\alpha$ smaller than unity. 
The drift velocity monotonically increases with force, though the scaling behavior varies considerably across different regimes of $F$. Moreover, the amplitude of the velocity oscillations displays rich behavior: it first undergoes a decay with force, reaches a minimum, and then presents a revival, increasing with force.

\end{abstract}

\date{\today}

\maketitle

\section{Introduction}

The out-of-equilibrium dynamics of quantum many-body systems has emerged as a central topic in modern physics. The motion of a distinguishable particle (an impurity) through an environment is an archetypal problem that governs the physics of a large class of systems \cite{landau+49,Devreese_2009,AnnalsKamenev}. 
Of particular interest are one-dimensional (1d) systems, where the interplay of correlations and restricted phase space can lead to fundamentally different physical phenomena relative to their higher-dimensional analogs \cite{Giamarchi}. In this work, we focus on a 1d system of bosons. While the dynamics of a repulsively interacting impurity has attracted a lot of attention \cite{ThBlochSolitons,AnnalsKamenev,PhysRevLett.108.207001,QFlutterNature,PhysRevE.90.032132,Meinert945,QuenchZvonarev, PhysRevLett.122.183001,MyPRLdissipative,PhysRevLett.130.220403,Will_2023,Quench, BlochAP,BlochSolitons,Scopa2026}, the regime of attractive impurity-boson interactions remains comparatively unexplored \cite{Ion1d,Attractive}. The latter exhibits a very interesting dynamics after a sudden injection of an impurity with a nonzero initial velocity.  Notably, a heavy and fast impurity interacting via a contact potential undergoes undamped transient long-lived velocity oscillations before reaching a stationary state  \cite{Attractive}. The underlying mechanism is the transitory trapping and oscillations of the depletion cloud in the vicinity of the density peak centred at the impurity position.

A peculiarity of 1d quantum liquids in the presence of an impurity is the periodicity of the ground-state energy as a function of momentum \cite{lamacraft2009dispersion}. In a system of weakly-interacting bosons, under the action of an infinitesimaly small driving force acting on a repulsively interacting impurity, the system follows its momentum dependent ground state in time \cite{AnnalsKamenev, BlochAP}. As a result, the average impurity velocity is zero, and the impurity oscillates around a fixed position in space, while the supercurrents carry the momentum pumped into the system. These are the Bloch oscillations, occurring in the absence of a periodic potential \footnote{See, for example, Refs.~\cite{Bruderer2008,PhysRevA.90.063610} for work on impurity Bloch oscillations in periodic potentials.}. Increasing the magnitude of the force, the impurity acquires a nonzero mean velocity and the center of oscillations drifts in time \cite{AnnalsKamenev}. Recently, we have shown that the drifted Bloch oscillations persist in a large interval of forces due to the periodic generation of nonlinear excitations of the host bosons that take away one part of the momentum transferred to the system while the impurity drift velocity remains constant in time \cite{BlochAP}.

In this work, we study the dynamics of an attractively interacting impurity driven by a constant external force through a system of weakly-interacting homogeneous bosons. In the absence of a force, the ground state takes a soliton-like form on both sides of the impurity with a constant shift
of its center and its phase. It describes bosons gathered around the impurity and forming a peak in the boson density that moves together with the impurity. Its energy is a periodic function of the total system momentum. However, this state exhibits sectors of forbidden momenta \cite{Attractive}. These sectors originate from the constraint that the stationary impurity velocity must be smaller than the critical velocity, thereby imposing specific conditions on the total system momentum. In contrast, the stationary state reached through the relaxation dynamics following a sudden quench of the impurity-boson interaction -- for an arbitrary system momentum -- is given by the state described above \cite{Attractive}. This is made possible by the momentum transfer from the impurity to the bath via the emission of excitations that propagate away from the impurity.
The present work studies the following key questions: Are there the Bloch oscillations? What is their underlying mechanism? What kind of a state is realised at forbidden momenta, especially in the limit of a very small force? How does the drift velocity, the Bloch amplitude and the time period of oscillations depend on the magnitude of the applied force?  How does the dynamics change if the impurity mass exceeds the critical one and the interval of forbidden moment disappears, accompanied by the emergence of metastable states? 

The paper is organized as follows. We present the model in Sec.~\ref{sec:model}. Next, in Sec.~\ref{sec:groundstate}, we study a stationary state of the system as a function of a total system momentum reached in the protocol where the impurity-boson interaction is switched on adiabatically slowly, in the absence of a driving force.  In Sec.~\ref{sec:Bloch}, we study the dynamics of both the impurity and the host bosons in the presence of an external driving force acting on the impurity. Section \ref{sec:critical} investigates the case of an impurity with a mass greater than the critical mass. We conclude with a summary of our main findings in Sec.~\ref{sec:conclusions}.

\section{Model\label{sec:model}}

We study the dynamics of a single impurity of mass $M$ through a one-dimensional bosonic environment at zero temperature. The impurity is subject to a constant external force $F$. The Hamiltonian of the system takes the form
\begin{align}
	\hat{H}=\frac{\hat{P}^2}{2M}+\hat{H}_b+G \hat{\Psi}^\da(\hat{X})\hat{\Psi}(\hat{X})-F \hat{X}.
	\label{eq:H}
\end{align}
The momentum and position operators of the impurity are denoted by $\hat{P}$ and $\hat X$, respectively. 
The contribution $\hat H_b$ describes bosons of mass $m$ with a repulsive contact interaction of strength $g$, and reads as
\begin{align}
\hat{H}_b=\int \mathrm{d}x\left[-\hat\Psi^\da(x)\dfrac{\hbar^2\partial_x^2}{2m}\hat\Psi(x)+\frac{g}{2}\hat\Psi^\da(x)\hat\Psi^\da(x)\hat\Psi(x)\hat\Psi(x)\right].
\end{align}
The bosonic single-particle operators $\hat\Psi^\da(x)$ and $\hat\Psi(x)$ satisfy the commutation relation $[ \hat\Psi(x) , \hat\Psi^\da(x')]=\delta(x-x')$. 
The third contribution in \eq{eq:H} models the impurity-boson interaction energy.  The impurity couples to the boson density at its position with a coupling constant $G$. We are interested in an attractive interaction, $G<0$.

In order to recover periodic boundary conditions, we apply a time-dependent unitary transformation $\hat{U}_1(t)=e^{i F \hat{X} t/\hbar}$ that acts as
$
\hat{H}_1(t)=\hat{U}_1^\da \hat{H}\hat{U}_1-i\hbar \hat{U}_1^\da \partial_t \hat{U}_1.
$
Then, we perform the Lee-Low-Pines transformation \cite{LeeLowPines} as $\hat{\mathcal{H}}=\hat{U}_2^\da \hat{H}_1\hat{U}_2$ where $\hat{U}_2=e^{-i \hat{X} \hat{p}_b/\hbar}$. Here, $\hat{p}_b=-i\hbar\int \mathrm{d}x  \hat\Psi^\da(x)\partial_x \hat\Psi(x)$ is the momentum of the host bosons.
The impurity is situated at the origin in the new reference frame and the Hamiltonian $\hat{\mathcal{H}}$ does not depend on $\hat{X}$. As a result, in the new referent system, $\hat{P}$ does not vary in time. Thus, we replace it by a number $p$. The total momentum of the system transforms is 
$\hat{U}_2^\da \hat{U}_1^\da (\hat{P}+\hat{p}_b)\hat{U}_1 \hat{U}_2=\hat{P}+Ft=p+Ft$.
The final Hamiltonian takes the form \cite{BlochAP}
\begin{align}\label{eq: HLee-Low}
	\hat{\mathcal{H}}(t)=\frac{(p+F t-\hat{p}_b)^2}{2M}+\hat{H}_b+G \hat{\Psi}^\da(0)\hat{\Psi}(0).
\end{align}
We are interested in the dynamics of the system with $p=0$, where both the bosons and the impurity are initially in their zero-momentum ground state and do not interact. Then, the impurity is injected into the system of bosons and starts experiencing the interaction potential of bosons and the driving force $F$.  
 
Henceforth, we are interested in weakly-interacting bosons. In this regime, the dimensionless parameter $\gamma=m g/\hbar^2 n_0$ satisfies $\gamma\ll 1$. Here, $n_0$ denotes the mean boson density. We thus employ the small-$\gamma$ expansion of the bosonic single-particle operator \cite{pitaevskii_bose-einstein_2003,sykes_drag_2009,CasimirNewJPhys}, and in the leading order in $\gamma$, we obtain the equation of motion for the condensate wave function $\Psi_0$ to be
\begin{align}
i\hbar \partial_t{{\Psi_0}(x,t)}=\Bigg[&-\frac{\hbar^2}{2}\left( \frac{1}{m}+\frac{1}{M}\right)\partial_x^2+g |{\Psi_0}(x,t)|^2\notag\\&+G\delta(x)+i\hbar V(t)\partial_x\Bigg] {\Psi_0}(x,t).
\label{eq:mean-field1}
\end{align}
Here, $V(t)$ denotes the impurity velocity in the laboratory frame. It is given by
\begin{align}\label{eq:Vimp}
V(t)=\frac{p+F t}{M}+i\frac{\hbar}{M}\int \mathrm{d}x  \Psi_0^*(x,t)\partial_x \Psi_0(x,t).
\end{align}

\section{Finite momentum ground state at $F=0$ \label{sec:groundstate}}

In this section, we consider a stationary solution of \eq{eq:mean-field1} with periodic boundary conditions in the absence of a force, for a total momentum $p$ of the system, after an adiabatically slow turn-on of the impurity-boson interaction. We assume that the impurity velocity takes a time-independent value $V$. The stationary solution can be evaluated analytically, for more details see the analysis of Ref.~\cite{Attractive}. Here, we consider only some of its properties that are relevant for the subsequent discussions.

In the reference frame co-moving with the impurity, the boson density reads as
\begin{align}
n(x)=n_0\left[a^2+b^2\tanh^2{\left(\frac{b|x|}{\xi\sqrt{1+m/M}} +x_0\right)}\right].
\label{densityStationary}
\end{align}
Here, $\xi=\hbar/\sqrt{m g n_0}$ is the healing length, while $v=\sqrt{g n_0/m}$ is the sound velocity. Additionally,  we have introduced $a={V}/{v\sqrt{1+m/M}}$ and $b=\sqrt{1-a^2}$. The shift parameter $x_0$ satisfies 
\begin{align}\label{eq:x0}
G(a^2+b^2 \tanh^2{x_0})=\hbar v b^3 \sqrt{1+m/M} \tanh{x_0}\; \textrm{sech}^2{x_0}. 
\end{align}
Note that \eq{eq:x0} admits solutions only for the impurity velocity smaller than the critical velocity $|V|\leq v_c$. The critical velocity is given by 
\begin{align}\label{eq:vc}
v_c=v\sqrt{1+m/M}.
\end{align}
This result is in agreement with the findings for an infinitely heavy impurity \cite{PhysRevA.66.013610}.
Moreover, there is just one physical solution of Eq.~(\ref{eq:x0}) describing an increase of the boson density at the impurity position, $n(0)> n_0$. This solution is a complex number $x_0$ with the imaginary part $i\pi/2$. It satisfies $\tanh{x_0}>1$ and thus $\tanh{\left(\beta |x| +x_0\right)}=\coth{\left[\beta |x| +\mathrm{Re}(x_0)\right]}$ for $\beta\in \mathbb{R}$. To conclude, the density profile (\ref{densityStationary}) actually describes a peak in the boson density centred at the impurity position. The other two solutions of Eq.~(\ref{eq:x0}) are studied in  Ref.~\cite{Attractive}.

The impurity also causes a phase increase $\theta$ across its position 
\begin{align}
\theta=&2\arctan\left(\frac{b }{a}\tanh{x_0}\right)-2 \arctan \left(\frac{b}{a} \right).
\label{eq:theta}
\end{align} 
In order to satisfy periodic boundary conditions, the contribution $-\theta x/L$ occurs in the phase of the condensate wave function. Here, $L$ denotes the length of the system.
As a result, the total system momentum $p$ as a function of stationary impurity velocity $V$ is given by
\begin{align}
p=&M{V} - {2\hbar n_0}ab \left(1 - \tanh{x_0} \right) -\hbar n_0 \theta + 2 k\pi\hbar n_0,
\label{eq:momentum}
\end{align}
for $\pi \hbar n_0 (2k-1) <p\leq\pi\hbar n_0(2 k+1)$. Here $k$ is an integer. Note that increasing $V$ from zero to $v_c$,  both the density peak at the impurity position and the phase variation across the impurity increase.

The difference of the ground state energy of the Hamiltonian (\ref{eq: HLee-Low}) at a given total momentum $p$ and its ground state energy at zero momentum in the absence of the impurity, is given by \cite{Attractive}
\begin{align}
\frac{E_{p}}{g n_0}={}&\frac{2}{3}b^3\frac{\sqrt{1+m/M} }{\sqrt{\gamma }}\left[1-\tanh^3(x_0)\right]+\frac{MV^2}{2g n_0}\notag\\ &+\frac{1}{3} b^3
\frac{\sqrt{1+m/M} }{\sqrt{\gamma }} \left[\tanh ^3(x_0)-3 \tanh(x_0)+2\right].
\label{eq:PolaronEnergy}
\end{align}
Here $V$ and $x_0$ satisfy Eqs.~(\ref{eq:x0}) and (\ref{eq:momentum}). Shifting the momentum $p\to p\pm 2\pi\hbar n_0$, due to the last term in \eq{eq:momentum}, both $V$ and $x_0$ remain unaltered.
Consequently, the energy (\ref{eq:PolaronEnergy}) is periodic in momentum with a period of $2\pi\hbar n_0$.
Actually this momentum is carried by supercurrents, and thus it results in zero change in energy.

\begin{figure}
    \centering
    \includegraphics[width=1\columnwidth]{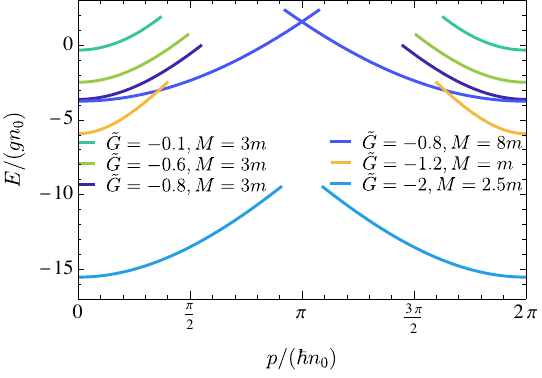}
   \caption{The dimensionless energy (\ref{eq:PolaronEnergy})  as a function of the momentum for different set of parameters. Here, $\gamma=0.1$.}
    \label{fig:groundstate}
\end{figure} 

Figure \ref{fig:groundstate}  shows the energy (\ref{eq:PolaronEnergy}) as a function of momentum for different values of the coupling constant and the impurity mass. We see that the physical stationary solution (\ref{densityStationary}) is allowed only in a certain momentum interval. This interval is determined by \eq{eq:momentum}, with the impurity velocity $V$ restricted to the interval $[-v_c,v_c]$. However, as the impurity mass or its coupling constant increases, this interval expands and the two energy branches approach one other (see Fig.~\ref{fig:groundstate}). As in the case of repulsive impurity-boson interaction \cite{PhysRevLett.108.207001}, at critical mass $M_c$, the two energy branches touch each other, resulting in cusps in the ground-state energy at momenta $p=(2m+1)\pi \hbar n_0$, with $m$ being an integer, and the interval of forbidden momenta disappears. At higher values of the impurity mass, $M>M_c$, the energy branches cross each other leading to metastable states.

In the following, we consider a weak to moderate coupling constant $\tilde{G}$ in order to avoid the collapse of bosons at the impurity position \cite{Attractive}.
	
\section{Dynamics under the drive\label{sec:Bloch}}

\begin{figure*}
    \centering
    \includegraphics[width=1\columnwidth,valign=b]{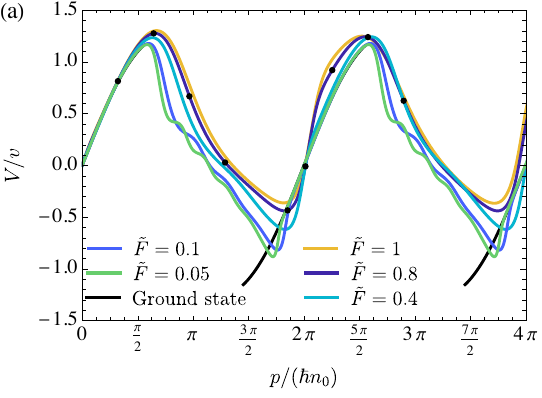} \phantom{aaa}
    \includegraphics[width=1\columnwidth]{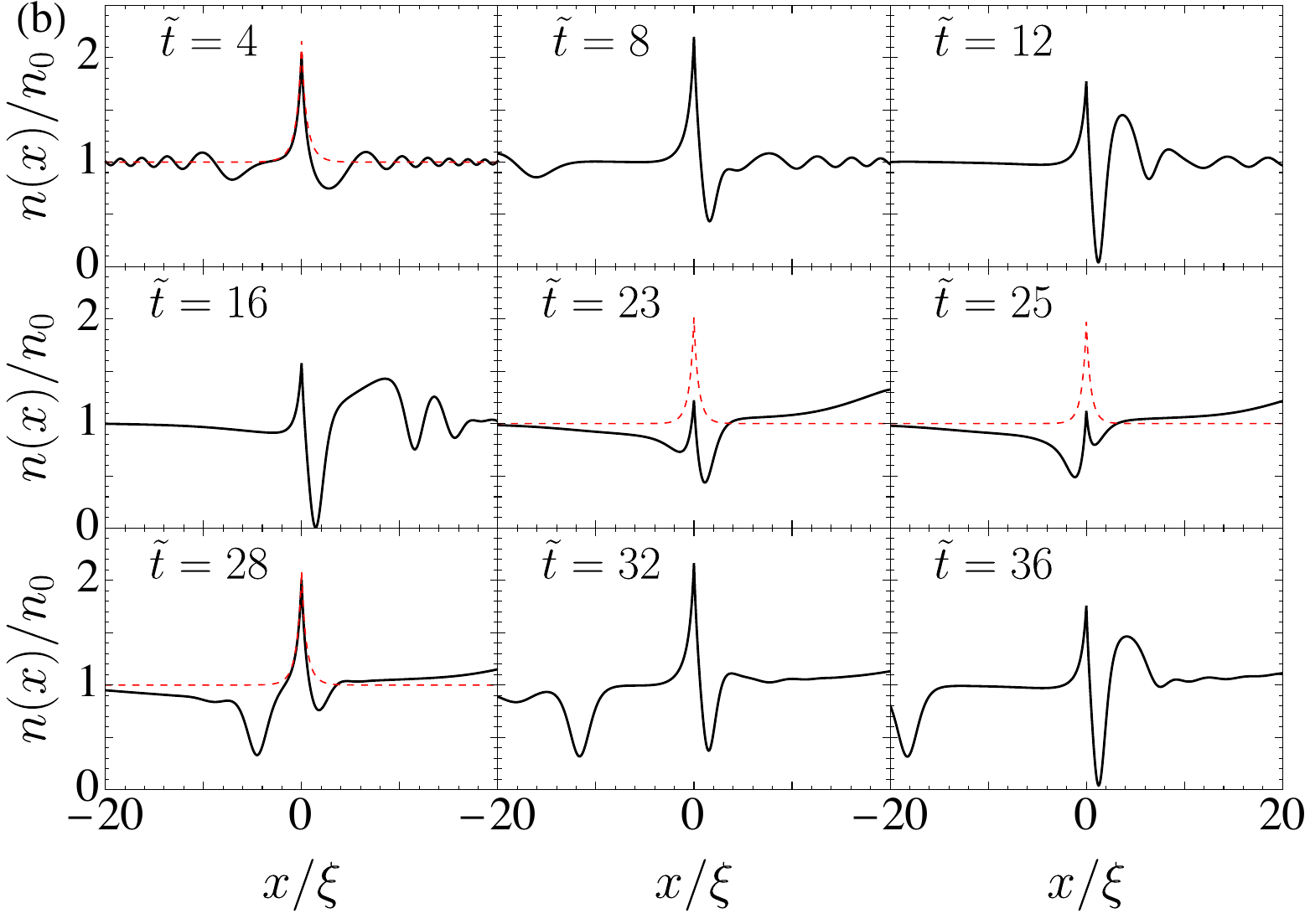}
    \caption{(a) Impurity velocity as a function of the total momentum of the system $p=\hbar n_0\sqrt{\gamma} \tilde{F}\tilde{t}$ for different values of dimensionless force $\tilde{F}$ for $\tilde{G}=-0.8$, $M=3m$, and $\gamma=0.1$. The velocity periodicity is established after the first oscillation for $\tilde{F}=0.8$ and $\tilde{F}=1$. The black curve shows the ground-state impurity velocity (\ref{eq:momentum}) as a function of $p$. The black points along the velocity curve for $\tilde{F}=0.8$ denote the time-stamps at which the densities in Fig.~(b) are displayed. (b) The time evolution of the density profile $n(x)/n_0$ for $\tilde{F}=0.8$ for the aforementioned parameters. The first $V_{\text{max}}$ is reached at $\tilde{t}=8.1$, the first $V=0$ at $\tilde{t}=16.3$, and the first $V_{\text{min}}$ at $\tilde{t}=22.7$. The second zero velocity occurs at $\tilde{t}=25$, while the second $V_{\text{max}}$ happens at $\tilde{t}=31.7$. Numerically calculated dimensionless period of the oscillations is $Tgn_0/\hbar = 23.85$. The red dashed line denotes the stationary configuration (\ref{densityStationary}) at $p=Ft$ where it exists.}
    \label{fig:V}
\end{figure*}

In this section, we study the system dynamics after a sudden switching on of the impurity-boson coupling and the driving force at $t=0$. Initially, both the impurity and the bosons are in their zero-momentum ground states. We numerically solve Eq.~(\ref{eq:mean-field1}) with periodic boundary conditions, and monitor the time-evolution of $\Psi_0(x,t)$ and $V(t)$ using an implicit conservative finite-difference scheme \cite{GPE_discretization}. The system size is taken long enough such that there are no boundary effects throughout the simulation.

In the absence of an external force, the impurity remains static. It triggers an emission of dispersive density shock waves that propagate symmetrically away from the impurity. Meanwhile, a peak in the bosons density forms at the impurity position following the analytical form (\ref{densityStationary}) for $V=0$. 
This picture completely changes in the presence of an external force. For a sufficiently small force, the impurity velocity periodically oscillates in time. The impurity velocity is characterized by a nonzero average value over its time period, that defines the impurity drift velocity. The latter is a time independent quantity. Thus, the impurity undergoes drifted Bloch oscillations. 

\subsection{Physical mechanism\label{sec:BlochMech}}

\begin{figure*}
    \centering
    \includegraphics[width=0.67\columnwidth]{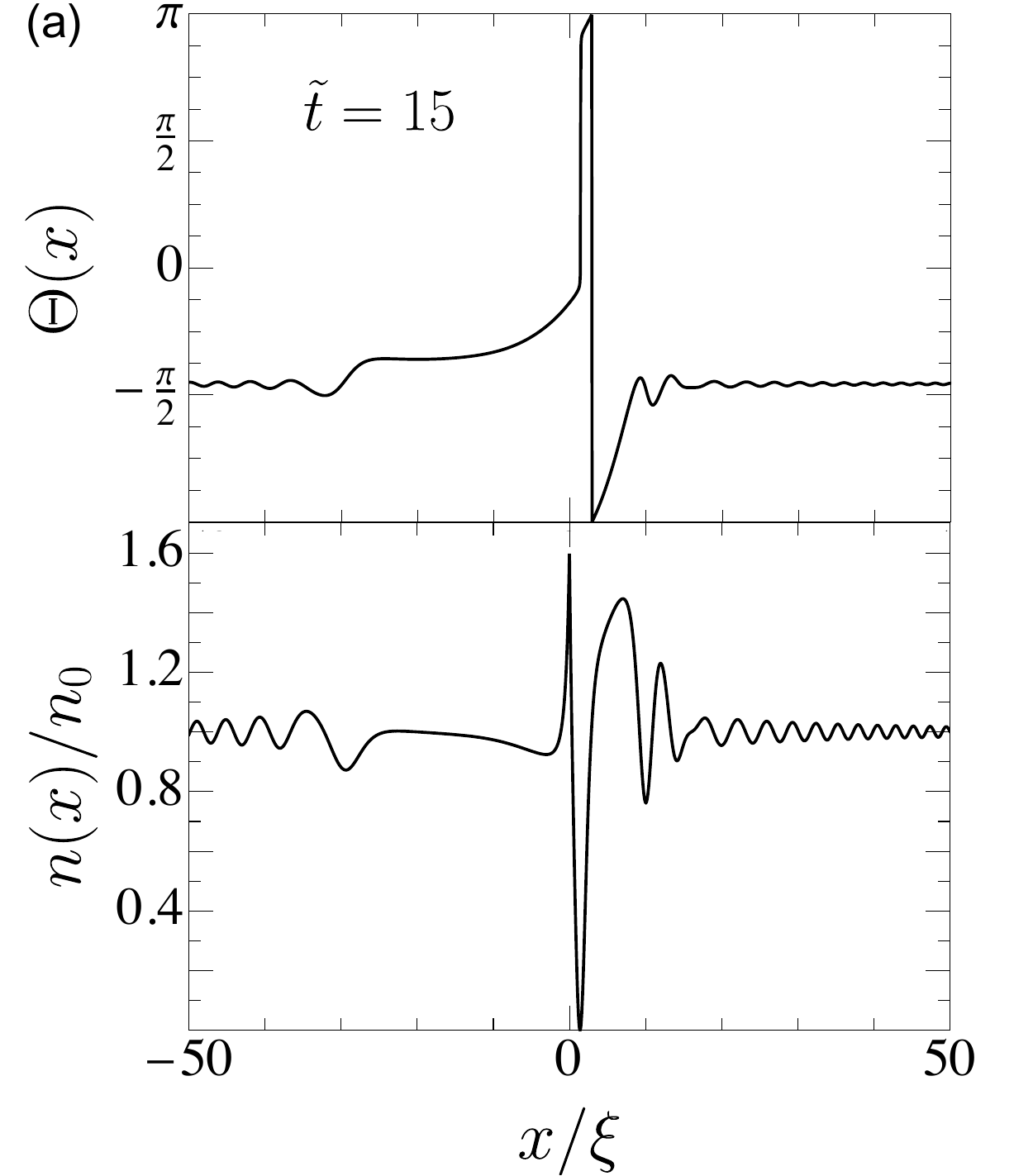}
    \includegraphics[width=0.67\columnwidth]{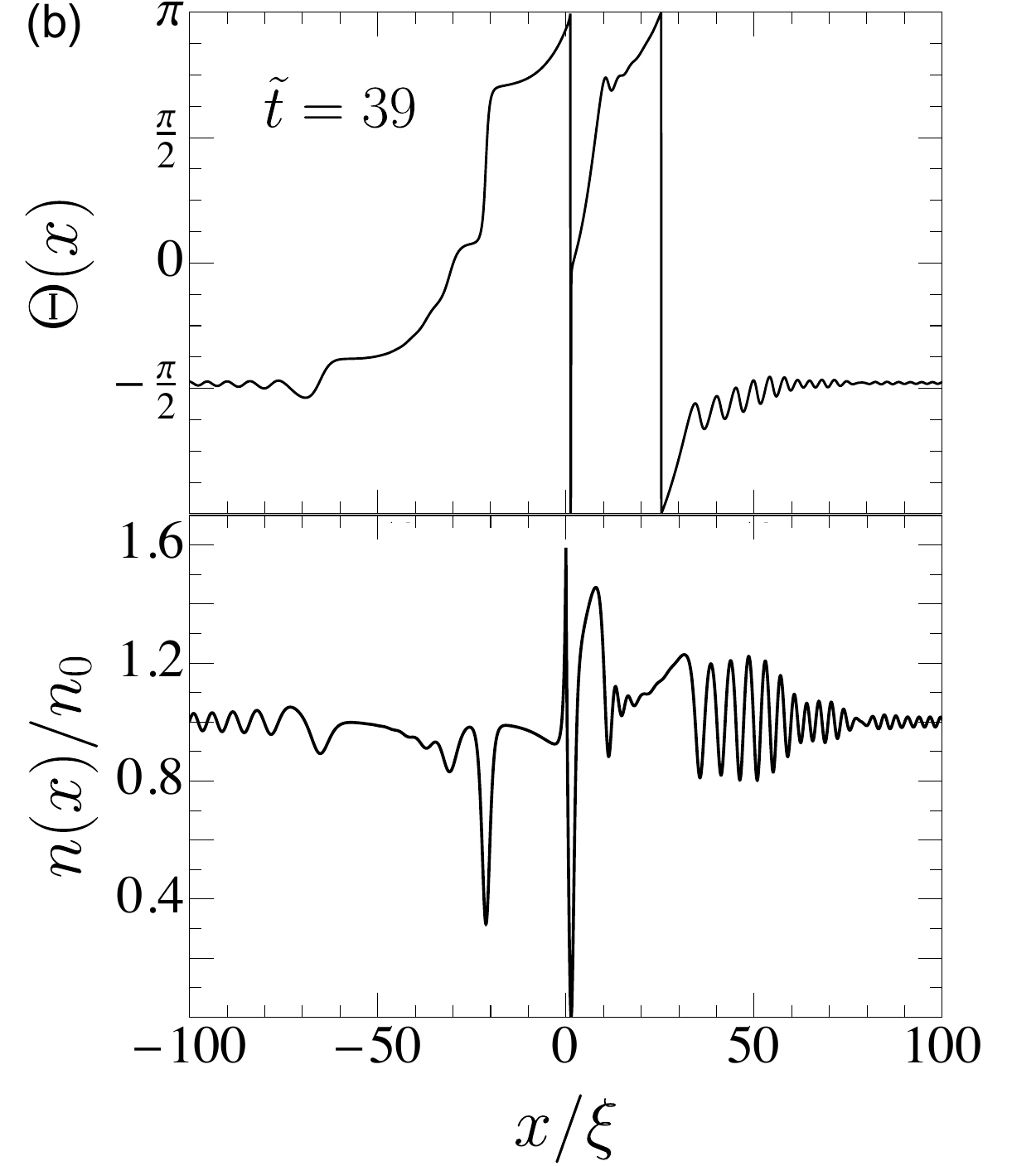}
    \includegraphics[width=0.67\columnwidth]{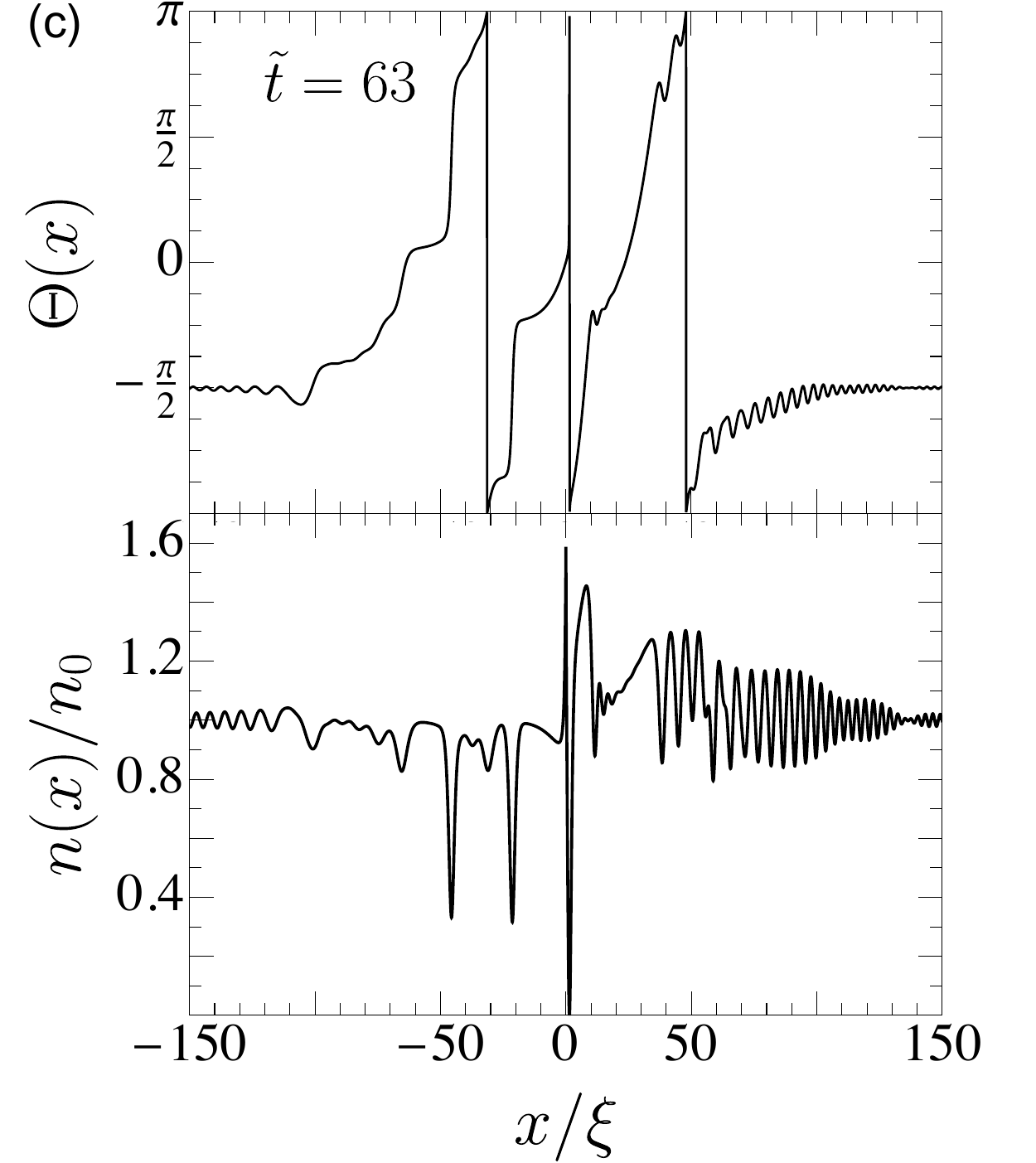}
    \caption{ Phase $\Theta(x)$ of the condensate wave-function at three time instances for $\tilde{G}=-0.8,M=3m,\tilde{F}=0.8,\gamma=0.1$. The dimensionless time period is $Tgn_0/\hbar=23.85$.}
    \label{fig:Theta}
\end{figure*}

We define the dimensionless impurity strength, force and time as
\begin{align}\label{eq:dimensionless}
\tilde{G}=G/\hbar v, \quad
\tilde{F}=F \xi/g n_0, \quad \tilde{t}=t g n_0/\hbar,
\end{align}
respectively. The drifted Bloch oscillations are illustrated in Fig.~\ref{fig:V}a, where we consider the case of $\tilde{G}=-0.8$, $M=3m$ and $\gamma=0.1$. We thus see that the shape of the oscillations and the drift velocity depend on the magnitude of the force. We will study their dependence on $F$ in more details below. First, we focus on the explanation of the underlying physical mechanism of the Bloch oscillations.
The total momentum of the system increases in time as $p=F t$, while the impurity velocity oscillates periodically in time. Thus, the momentum pumped into the system over one period is actually transferred to the host boson. 

We now explain the dynamic of the bosons, the periodically emitted excitations and the momentum carried by the gradients of the phase. 
First, we present the time evolution of the boson density for aforementioned parameters and $\tilde{F}=0.8$ in Fig.~\ref{fig:V}b. The corresponding velocity temporal dependence is shown in Fig.~\ref{fig:V}a.
As the impurity coupling is turned on, the bosons gather around the impurity forming a peak at the impurity location, while the first set of dispersive density shock-waves is emitted. Under the influence of the driving force, the impurity accelerates, resulting in an increase in the bosonic peak height. Simultaneously, a depletion hole forms in front of the impurity (Fig.~\ref{fig:V}b). The impurity velocity reaches its maximal value at $\tilde{t}=8.1$. Then, the impurity slows down by imparting the excessive momentum into the condensate in the form of a dispersive shockwave that is now released in front of the impurity. At $\tilde{t}=12$, one can observe in Fig.~\ref{fig:V}b the density wave front being built in front of the impurity. This shockwave is fully released before the impurity velocity vanishes at $\tilde{t}=16.3$. As the impurity velocity decreases from its maximal value to zero, the density peak decreases, while the depth of the depletion region situated in front of the impurity increases leading to a complete depletion visible at $\tilde{t}=16$. 
The impurity changes its direction of motion at  $\tilde{t}=16.3$ and starts accelerating in the opposite direction. 
Notice that now the density peak decreases in time, despite the acceleration, contrary to the intuition gained from the stationary solution (\ref{densityStationary}). However, this solution is not expected to be realized in this interval of the momenta anyway. \begin{figure*}
    \centering
    \includegraphics[width=2.1\columnwidth, clip]{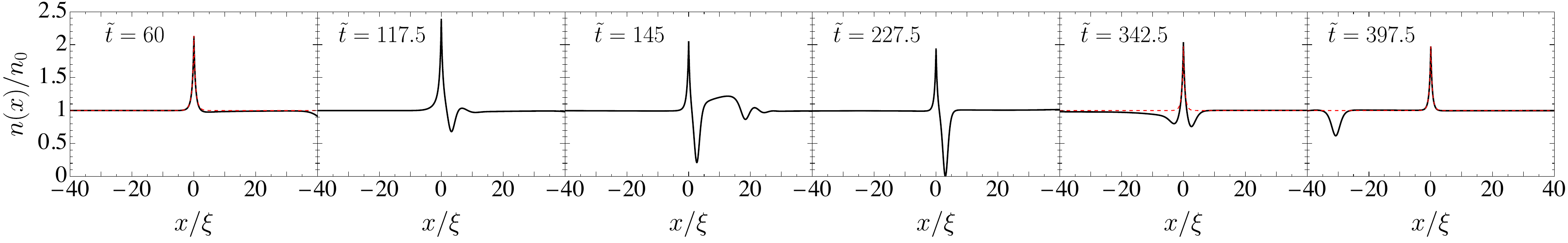}
    \caption{The time evolution of the boson density profile $n(x)/n_0$ over the first period of oscillation for $\tilde{G}=-0.8$, $M=3m$, $\tilde{F}=0.05$, and $\gamma=0.1$. The corresponding velocity curve is shown in Fig.~\ref{fig:V}. The impurity velocity is maximal at $\tilde{t}=116.3$, while $\tilde{t}=145$ shows the emission of the density shockwave.  $V=0$ at $\tilde{t}=227$, $V=V_{\text{min}}$ at $\tilde{t}=342.5$, and $\tilde{t}=397.5$ the velocity reaches zero again. The red dashed plots represents the analytical stationary configurations (\ref{densityStationary}) at $p=Ft$ when they exist.
    }
    \label{fig:smallF}
\end{figure*}

\begin{figure}
	\centering
	\includegraphics[width=1\columnwidth,valign=t]{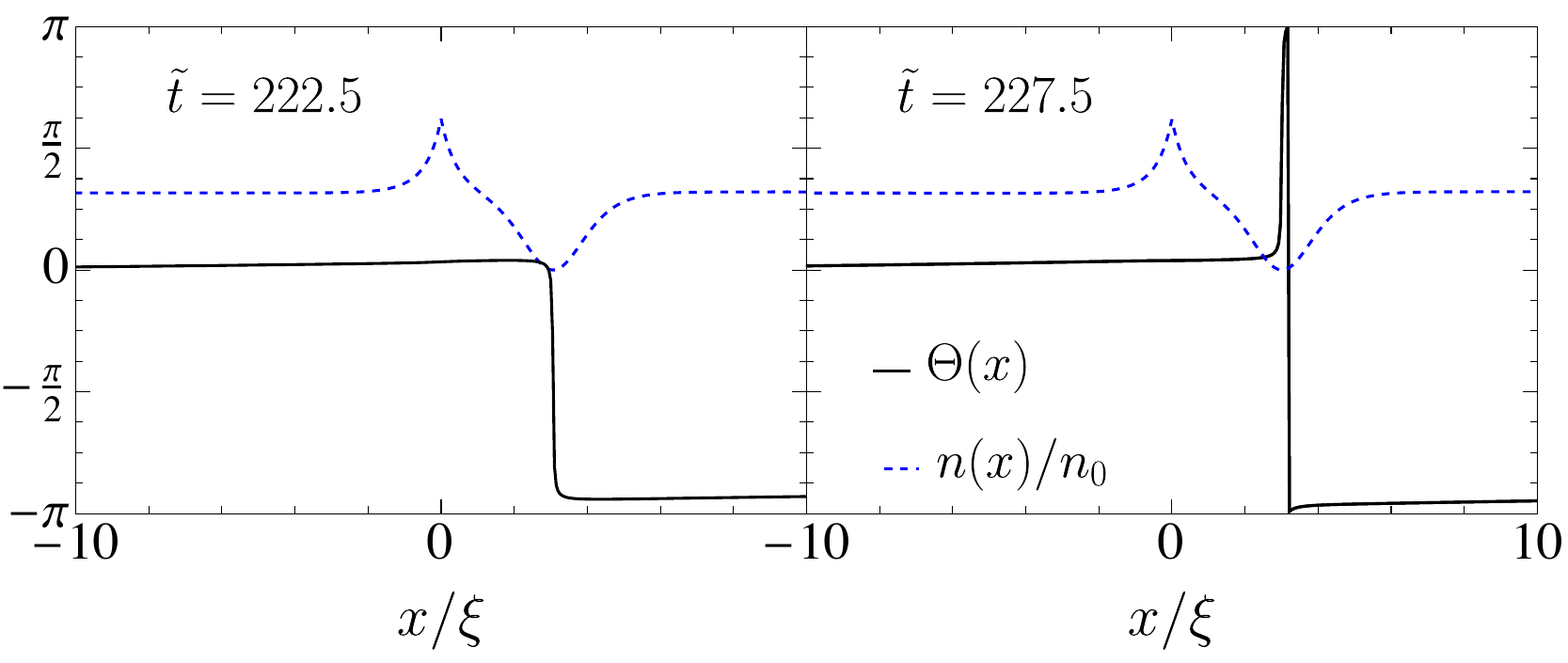}
	\caption{The time evolution of the boson density (blue dashed line) and phase profile (black solid line) in the vicinity of the impurity for $\tilde{G}=-0.8$, $M=3m$, $\tilde{F}=0.05$, and $\gamma=0.1$. The impurity velocity is $V=0$ at $\tilde{t}=227$.}
	\label{fig:PSlip}
\end{figure} 
The previously formed depletion cloud, now located behind the impurity, decreases in depth, while a new depletion hole is forming in front of the impurity. The impurity velocity reaches its minimum at $\tilde{t}=22.7$, and starts decelerating rapidly, while the depletion depth in front of the impurity exhibits a continuous increase. The impurity arrives at zero velocity at $\tilde{t}=25$ and changes the direction of motion. Its acceleration is now accompanied by the emission of a grey soliton behind the impurity, see Fig.~\ref{fig:V}b. Meanwhile, the density peak increases. At $\tilde{t}=31$, the velocity reaches the maximal value, while the soliton continues its propagation away from the impurity. The emission of the solitons and density waves is periodically repeated, and ensures the periodicity of the impurity velocity in time.

Next, we study the time evolution of the phase of the condensate wave function. It is shown in the top panels of Fig.~\ref{fig:Theta} for the aforementioned parameters, over three periods of the oscillations. The phase variation is taking place in the region between the firstly emitted density shock waves moving in the opposite directions. The phase gradients are the most pronounced across the solitons and the bosonic peak at the impurity position, but we distinguish also additional contributions not carried by these structures. Importantly, the total phase variation over the system length increases by $2\pi$ over each time period. This is represented by an additional vertical line appearing within one time period in the phase profile in Fig.~\ref{fig:Theta}, since the phase is defined in the interval $[-\pi, \pi)$. The lower panels of Fig.~\ref{fig:Theta} display the corresponding density profile, showing the periodic emission of solitons and shock waves, as well as the interferences of the shock waves emitted at different times (Fig.~\ref{fig:Theta}c). 

Note that also in the case of a repulsive impurity-boson interactions, the phase variation increases by $2\pi$ over a period \cite{AnnalsKamenev,BlochAP}. This increase can be understood from its finite momentum ground-state solution. However, for $G<0$, the underlying mechanism does not follow from the ground-state solution considered in Sec.~\ref{sec:groundstate}. 
We point out that the key element is the progressive formation of a total depletion hole in the vicinity of the density peak. This occurs as the total momentum increases within the interval of forbidden momenta, where the impurity velocity decays from its maximum value to zero.
The complete depletion is visible at $\tilde{t}=16$ in Fig.~\ref{fig:V}b. Note that this mechanism holds even in the limit of weak applied force.  This is illustrated in Fig.~\ref{fig:smallF} at $\tilde{F}=0.05$. The depletion becomes total near $p=\pi \hbar n_0$ for a very weak force.  Thus, it imposes locally the phase drop $\pi$ across its position (Fig.~\ref{fig:PSlip}). The latter is compensated by the rest of the system in order to sustain periodic boundary conditions. At shown times, the impurity velocity is nearly zero, and thus the phase drop across its position vanishes. Then, the impurity velocity crosses zero value at $\tilde{t}=227$, and the phase drop across the depletion hole attached to it changes sign, leading to a total phase variation of $2\pi$ along the system size (Fig.~\ref{fig:PSlip}). This process repeats over each time period.

\subsection{Weak-force regime}

Note that for $\tilde{F}=0.8$ the density peak around the impurity remains very different from the stationary configuration (\ref{densityStationary}) corresponding to the total momentum $p=Ft$ where this solution exists. The latter is shown by the red dashed lines in Fig.~\ref{fig:V}b. Also, for $\tilde{F}=0.8$, the phase increase of $2\pi$ over the system length takes place somewhat before the impurity velocity actually vanishes (Fig.~\ref{fig:Theta}). However, at smaller magnitudes of force, after reaching the minimal velocity the local density around the impurity is much better approximated by the stationary polaron (\ref{densityStationary}) at $p=Ft$. This is illustrated in Fig.~\ref{fig:smallF} for $\tilde{F}=0.05$. 
Moreover, this tendency is also visible in the small-$F$ regime of Fig.~\ref{fig:V}a, where the velocity curves shows a better overlap with the ground-state curve (\ref{eq:momentum}) as $F$ decreases. However, we stress that even at this very small force, $\tilde{F}=0.05$, the system does not follow the state (\ref{densityStationary}) after entering the allowed momenta range where the velocity decays to its minimal value (Fig.~\ref{fig:V}a). Moreover, the deviation is quite important in this interval.
\begin{figure}
	\centering
	\includegraphics[width=1\columnwidth,valign=t]{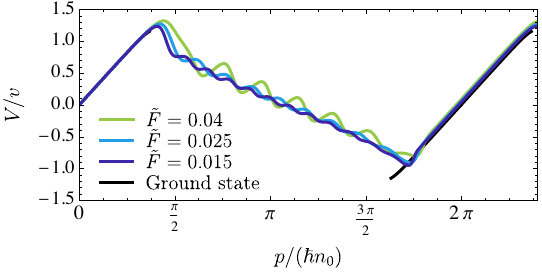}
	\caption{Impurity velocity as a function of the total momentum of the system $p=\hbar n_0\sqrt{\gamma} \tilde{F}\tilde{t}$ for different values of dimensionless force $\tilde{F}$ for $\tilde{G}=-0.1$, $M=3m$, and $\gamma=0.1$.}
	\label{fig:Suboscilations}
\end{figure} 

\begin{figure}
    \centering
    \includegraphics[width=1\columnwidth]{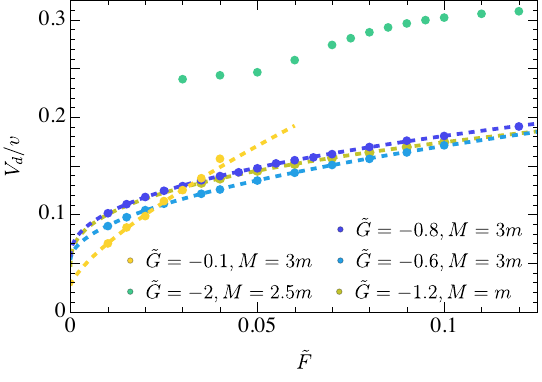}
   \caption{Drift velocity as a function of force in the small-force regime for different sets of parameters.  Fitting the data in the whole shown interval of forces, we get for $\tilde{G}=-0.8$ the dependence $0.34\tilde{F}^{0.44}+0.056$, for $\tilde{G}=-0.6$ it is $0.39 \tilde{F}^{0.52}+0.053$, and for $\tilde{G}=-1.2$ it reads as $0.30\tilde{F}^{0.37}+0.046$. For $\tilde{G}=-0.1$ we get $1.32 \tilde{F}^{0.74} + 0.027$ by omitting the largest value of the force, $\tilde{F}=0.04$, from the fit. These fitting functions are shown by the dashed lines.}
    \label{fig:VdSmallF}
\end{figure}

In the interval of forbidden momenta, in the regime of small forces, the impurity velocity first decreases sharply from $V_{\text{max}}$ and then undergoes small-amplitude superimposed oscillations as it decays towards $V_{\text{min}}$ (Fig.~\ref{fig:V}a). The number of these oscillations increases as $F$ decreases. This effect is more pronounced for a weak impurity-boson interaction, see Fig.~\ref{fig:Suboscilations}. The latter shows the impurity velocity for $\tilde{G}=-0.1$ while the impurity mass and the strength of the boson-boson interaction are the same as in Fig.~\ref{fig:V}a. Figure \ref{fig:Suboscilations} reveals that the amplitude of the superimposed oscillations and the sharp drop of the velocity from $V_{\text{max}}$ decay, while the number of oscillations increases as $F$ decreases. In the boson density, these oscillations manifest as oscillations of the density peak centred at the impurity position.

Moreover, the obtained results suggest that the state composed of the boson density peak situated at the impurity position and followed by a complete depletion cloud occurs also in the limit of a vanishingly small force in the interval of forbidden momenta at $V=0$. The density profile of this state is not symmetric under parity, and thus could lead to a nonzero drift velocity in the limit $F\to 0^+$. Note that the energy pumped into the system per oscillation period $T$ is $F\int _0^T V(t) dt=F V_d T$, where $V_d$ denotes the impurity drift velocity. This is the energy transferred to the host bosons during one period in the form of excitations. We find that the period of oscillation is $T=2\pi\hbar n_0/F$  at small magnitudes of the force. Thus, if a nonzero drift velocity is realized at $F\to 0^+$, then the energy of the emitted excitations per period saturates into a nonzero value. However, we observe that the emitted solitons become less profound and thus less energetic as the force decreases. The amplitude of the density waves also decreases with $F$. These effects are also visible in Figs.~\ref{fig:V}b and \ref{fig:smallF}. If the drift velocity becomes zero at $F\to 0$, then there are no emitted excitations and the entire momentum pumped into the system per period is carried by the constant phase gradients taking place over the system length and giving rise to supercurrents.

\begin{figure*}
    \centering
    \includegraphics[width=0.66\columnwidth]{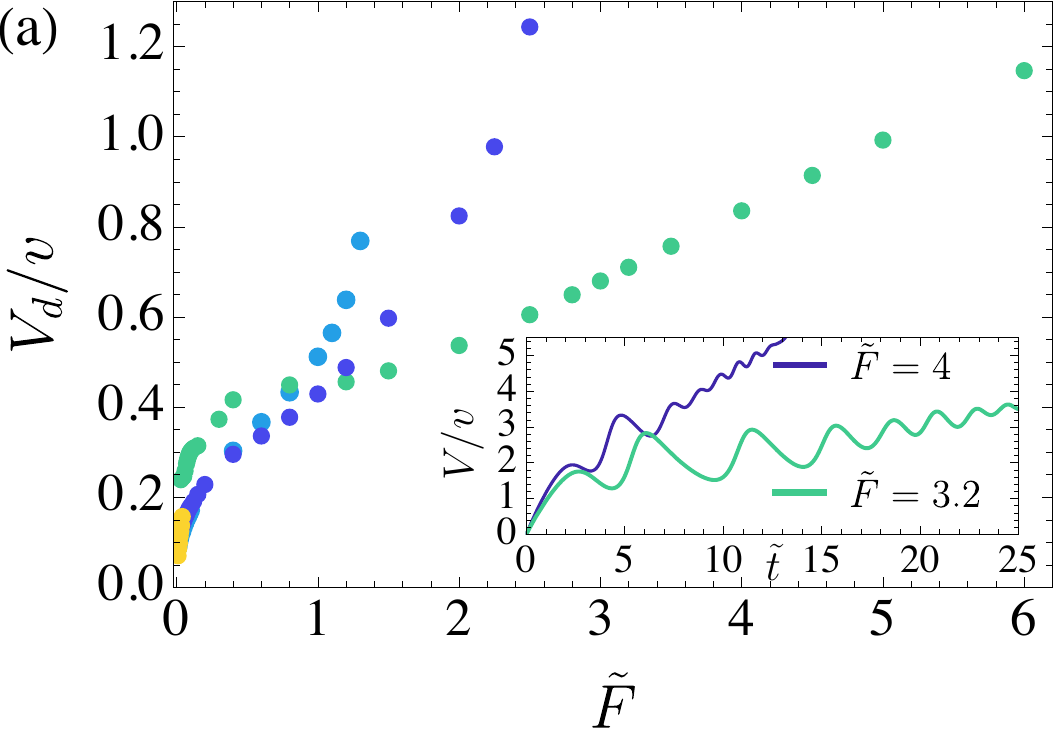}\phantom{a}
    \includegraphics[width=0.65\columnwidth]{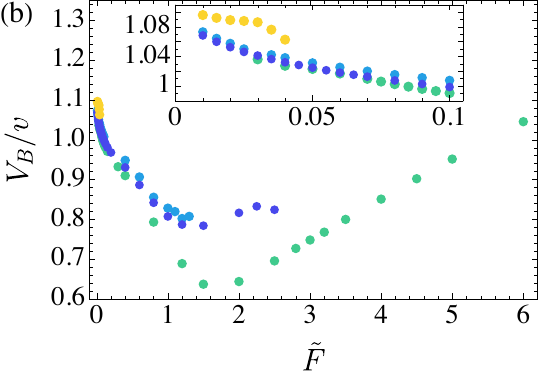}\phantom{a}
    \includegraphics[width=0.67\columnwidth]{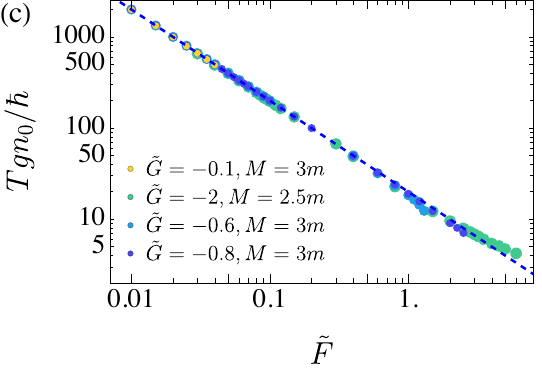}\phantom{a}
\caption{({a}) Dimensionless drift velocity $V_d/v$, ({b}) Bloch amplitude $V_B/v$, and ({c}) time period $Tgn_0/\hbar$ as a function of dimensionless force $\tilde{F}$ for different set of parameters.
The legend is shown in the panel (c).  The inset in the panel (a) shows $V/v$ as a function of $\tilde{t}$ for $\tilde{G}=-0.8$ and $M=3m$ at $\tilde{F}=3.2$ and $4$. The inset in the panel (b) shows $V_d/v$ for $\tilde{F} \leq 0.1$. In the panel (c), the fitting function is $Tgn_0/\hbar=2\pi/(\tilde{F}\sqrt{\gamma})$. Everywhere $\gamma=0.1$. }
    \label{fig:VdF}
\end{figure*}

Surprisingly, we find that in the small-$F$ regime, the drift velocity $V_d$ exhibits a sub-linear dependence on the force. 
Figure \ref{fig:VdSmallF} shows $V_d$ as a function of $F$ for different values of $\tilde{G}$ and $M/m$. In particular, fitting the obtained data with the form $V_d/v=\beta \tilde{F}^{\alpha}+V_{d0}/v$, it appears that $\alpha$ is non-universal and that $V_{d0}\neq 0$. However,  we cannot access the limit of an adiabatically small force numerically and give $\alpha$ and $V_{d0}$ in that limit. The values of the mentioned parameters depend on the interval of the considered values of the force. Moreover, the obtained values of $V_{d0}$ are very small. More precisely, fitting the data in the entire interval of forces shown in Fig.~\ref{fig:VdSmallF}, we get for $\tilde{G}=-0.8$ and $M=3m$ the dependence $0.34\tilde{F}^{0.44}+0.056$, for $\tilde{G}=-0.6$ and $M=3m$ it is $0.39 \tilde{F}^{0.52}+0.053$, and for $\tilde{G}=-1.2$ and $M=m$ it reads as $0.30\tilde{F}^{0.37}+0.046$. Note that for $\tilde{G}=-0.1$ and $M=3m$, an unlimited acceleration takes place already at 
$\tilde{F}=0.05$. Thus, we have omitted the largest value of the force $\tilde{F}=0.04$ from the fit. The function takes the form $1.32 \tilde{F}^{0.74} + 0.027$.  For the discussed sets of parameters, decreasing the absolute value of the impurity coupling constant $\tilde{G}$, while keeping fixed the impurity mass, results in an increasing exponent $\alpha$. If this behavior remains valid in the limit $F\to 0^+$,  one would expect $\alpha$ to tend to unity as $G\to 0^-$.  This is consistent with the physical picture that at arbitrary small and positive $G$ the exponent $\alpha$ becomes unity \cite{AnnalsKamenev, BlochAP}.
To conclude, the obtained results suggest that the impurity mobility $\sigma=V_d/F$ diverges as $F\to 0$. This is the case even if $V_{d0}$ becomes zero in the limit $F\to 0^+$, as expected. 

Peculiarly, at $\tilde{G}=-2$, the drift velocity exhibits a different behavior with respect to other parameters shown in  Fig.~\ref{fig:VdSmallF}. The drift velocity is considerably higher. Moreover, it undergoes a change in convexity, and exhibits a tendency of saturation into a non-zero value in the small-$F$ region (see Fig.~\ref{fig:VdSmallF}). We point out that the proximity to the critical mass  (see Fig.~\ref{fig:groundstate}) is responsible for these differences. We discuss these features further in Sec.~\ref{sec:critical} that is devoted to the case $M\geq M_c$. The ground-state energy dispersion (\ref{eq:PolaronEnergy}) for all the sets of parameters shown in Fig.~\ref{fig:VdSmallF} is displayed in Fig.~\ref{fig:groundstate}.

The amplitude of the impurity velocity oscillations $V_B$ decays with force in the small-$F$ regime, see the inset of Fig.~\ref{fig:VdF}b. Moreover, Fig.~\ref{fig:VdF}b suggests that keeping the magnitude of the force and the impurity mass fixed, $V_B$ decreases with $|\tilde{G}|$. 

If the system follows the state (\ref{densityStationary}) in the limit $F\to 0^+$ and the impurity does not overcome $v_c$ (\ref{eq:vc}) in the interval of forbidden momenta, then $V_B$ is given by $v_c$ in that limit. However, we find that in the region where the impurity velocity decays towards its minimal value and the system is expected to cross into the soliton-like state (\ref{densityStationary}), important deviations from this state are present even at very small forces (Figs.~\ref{fig:V} and \ref{fig:Suboscilations}). As a result, the maximal value of $V_B$ that we have reached numerically is $5\%$ lower than the critical velocity for $\tilde{G}=-0.1$. For $\tilde{G}=-0.6$ and $\tilde{G}=-0.8$, $V_{B,\mathrm{max}}$ is $7.1\%$  and $7.5\%$ lower  than $v_c$, respectively. For $\tilde{G}=-2$,  it is $12\%$ lower than $v_c$. We remind the reader that $v_c$ is $\tilde{G}$-independent.

\subsection{Dynamics as a function of the applied force}

Having discussed the small-force regime, in this section, we study  the system dynamics in a wide range of forces. Figure \ref{fig:V}a shows how the increasing magnitude of the force alters the shape of the oscillations, while the deviation of the impurity velocity from Eq.~(\ref{eq:momentum}) increases. The superimposed small-amplitude oscillations, present in the very weak force regime, disappear. The velocity curve gets shifted towards higher velocities, leading to an increase of both the maximal and the minimal velocities. Thus, the drift velocity increases with force. Moreover, the amplitude of the Bloch oscillations decreases in the range of forces shown in Fig.~\ref{fig:V}a, as well as the period of oscillations. Furthermore, at higher magnitudes of the force, the impurity oscillations become periodic after a longer time. For example, the periodicity is established after the first oscillation for $\tilde{F}=0.8$ and $\tilde{F}=1$ shown in Fig.~\ref{fig:V}a.

We stress that the Bloch oscillations cease to exist at sufficiently high values of force, where the impurity undergoes an unlimited acceleration in time. This dynamical regime is illustrated in the inset of Fig.~\ref{fig:VdF}a for $\tilde{G}=-0.8$. The value of the maximal force $F_{\text{max}}$ under which the Bloch oscillations are maintained increases with the absolute value of the impurity-bath coupling. For $M=3m$, we find that $\tilde{F}_{\text{max}} \in [3,3.2)$ for $\tilde{G}=-0.8$, $\tilde{F}_{\text{max}} \in [1.3,1.5)$ for $\tilde{G}=-0.6$, and $\tilde{F}_{\text{max}} \in [0.04,0.05)$ for $\tilde{G}=-0.1$. For $\tilde{G}=-2$, the oscillations persist in a very wide range of forces.

The impurity drift velocity $V_d$ is a monotonically increasing function of the force, see Fig.~\ref{fig:VdF}a. The corresponding small-$F$ regime is shown in Fig.~\ref{fig:VdSmallF}. In the large-$F$ regime, $V_d$ undergoes a faster increase with force at weaker impurity-boson interaction.
For $\tilde{G}=-2$, the drift velocity exhibits a rich behavior. After changing its convexity twice at smaller-$F$ values, the drift velocity increases linearly in the large-$F$ regime, see Figs.~\ref{fig:VdF}a and \ref{fig:VdSmallF}. 
We point out that the linear dependence takes place in a wide interval of forces.

Next, we study the behavior of the amplitude of the impurity velocity oscillations, $V_B$, as a function of force. 
We find that in a general case $V_B$ is a non-monotonic function of $F$, see Fig.~\ref{fig:VdF}b.
At small and intermediate values of $F$, $V_B$ decreases, first rapidly and then with a somewhat slower rate.
Then, $V_B$ reaches a local minimum and starts increasing with $F$. At stronger impurity-boson interaction, the interval of forces where $V_B$ increase becomes wider. Figure \ref{fig:VdF}b also suggests that the value of minimal $V_B$ decreases with $|\tilde{G}|$. In the wide interval of large values of $F$ at $\tilde{G}=-2$, $V_B$ increases nearly linearly with force. At $\tilde{G}=-0.1$, the interval of forces where the Bloch oscillations exist shrinks down considerably, and $V_B$ shows different behavior for considered forces. Contrary to the other data, the initial slope with $F$ is lower, and subsequently steepens. Moreover, $V_B$ is a monotonically decreasing function of $F$.

Finally, the behavior of the time period $T$ of the oscillations is shown in Fig.~\ref{fig:VdF}c. 
It is compared against the prediction $T = 2\pi \hbar n_0/F$  shown by the dashed line. In dimensionless units, the prediction takes the form $Tgn_0/\hbar = 2\pi/(\tilde{F} \sqrt{\gamma})$. 
For $\tilde{G}=-0.8$, the obtained results match quite well the dashed curve at small forces, $\tilde{F}\leq 0.4$, with less than a half percent of the relative error. For higher values of force, the deviation monotonically increases from $4\%$ at $\tilde{F}=0.6$ to $13\%$ for $\tilde{F}=2.5$. We stress that the obtained period is shorter than the expected one. Note that in the vicinity of the break-down of the Bloch oscillations, where the deviation of $T$ become important, the above explained dynamics of the phase increase over the system length per period does not hold anymore.
Similar behavior of the time period with $F$ is observed for other sets of parameters shown in Fig.~\ref{fig:VdF}c, apart from the case of $\tilde{G}=-2$. There, the oscillations exist in a very wide interval of forces, and the obtained period is longer than the expected one in the large-$F$ regime, while it is somewhat shorter than the expected one in the intermediate-$F$ region. 
 \begin{figure}
    \centering
    \includegraphics[width=1\columnwidth,valign=t]{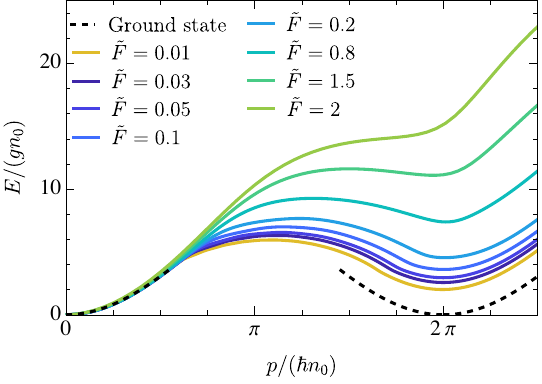}
    \caption{Total energy $E(\tilde{t})/(gn_0) = \tilde{F} \int_{0}^{\tilde{t}} (V(\tilde{t}')/v) d \tilde{t}'$ pumped into the system as a function of total system momentum $p(\tilde{t})/(\hbar n_0)= \sqrt{\gamma}\tilde{F} \tilde{t}$ for different values of force is compared against \eq{eq:PolaronEnergy}  shown by the dashed line. The parameters are $\tilde{G}=-0.8$, $M=3m$, $\gamma=0.1$.}
    \label{fig:Energy}
\end{figure}

Figure \ref{fig:Energy} shows the total energy pumped into the system by the external force, $E(t) = F \int_0^t V(t) dt$, as a function of the momentum $p=F t$ for different values of the force. Interestingly, initially, the system follows the ground state even at large magnitudes of the force. This behavior is also visible in the velocity dependence on $p$ displayed in Fig.~\ref{fig:V}. At later times, the system moves away from the ground-state energy dispersion (\ref{eq:PolaronEnergy}) shown by the black dashed line. The excess energy is transmitted to the nonlinear excitations, density waves and solitons. This energy increases with $F$, and thus also the drift velocity.

\section{Dynamics for $M\geq M_c$ \label{sec:critical}}

\begin{figure}
    \centering
    \includegraphics[width=1\columnwidth]{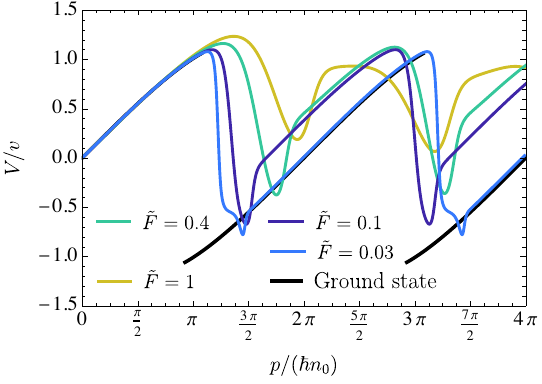}
    \caption{Impurity velocity as a function of the total momentum of the system $p=\hbar n_0\sqrt{\gamma} \tilde{F}\tilde{t}$ for different values of dimensionless forces $\tilde{F}$ for $\tilde{G}=-0.8$, $M=8m$, and $\gamma=0.1$.}
    \label{fig:V8m}
\end{figure}

\begin{figure*}
    \centering
    \includegraphics[width=0.66\columnwidth]{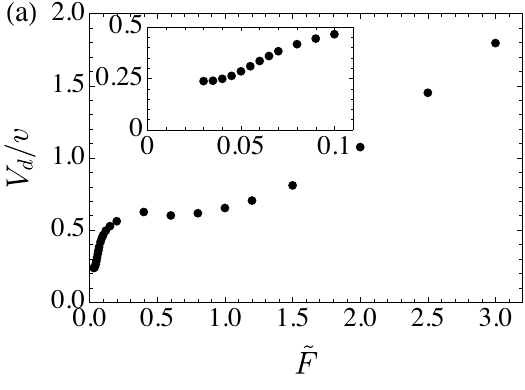}\phantom{a}
    \includegraphics[width=0.66\columnwidth]{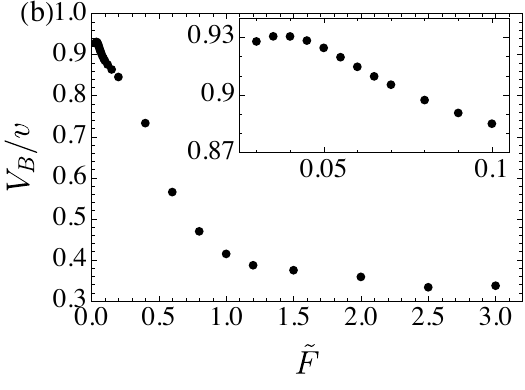}\phantom{a}
    \includegraphics[width=0.66\columnwidth]{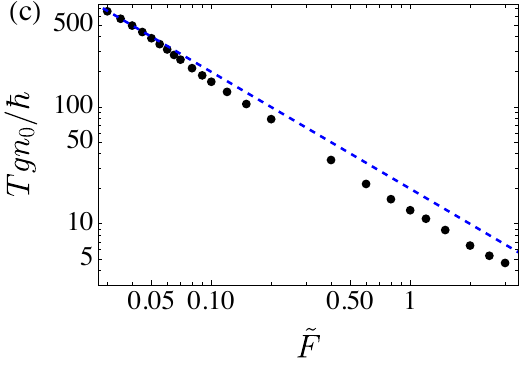}\phantom{a}
    \caption{(\textit{a}) Dimensionless drift velocity $V_d/v$, (\textit{b}) Bloch amplitude $V_B/v$, and (\textit{c}) time period $Tgn_0/\hbar$ as a function of dimensionless force $\tilde{F}$ for $\tilde{G}=-0.8$, $\gamma=0.1$ with $M=8m$. The insets in the panels (a) and (b) show the small force behavior of $V_d$ and $V_B$, respectively. In the panel (c), the dashed line represents $Tgn_0/\hbar = 2\pi/(\tilde{F}\sqrt{\gamma})$.}
    \label{fig:VDVB8m}
\end{figure*}

In this section, we study the dynamics of an impurity with a mass exceeding  the critical one.  As explained in Sec.~\ref{sec:groundstate}, in this case, the intervals of forbidden momenta disappear and the energy-momentum dispersion of the ground state becomes fundamentally different, exhibiting cusp-like singularities at $p = (2n+1)\pi \hbar n_0$ with $n \in \mathbb{Z}$ and metastable branches. This behavior is illustrated in Fig.~\ref{fig:groundstate} for $\tilde{G}=-0.8$ and $M=8m$.

In this regime, the drifted Bloch oscillations occur over a broad range of applied forces  as well. The impurity velocity oscillations are shown in Fig.~\ref{fig:V8m} for the above-mentioned parameters. 
We observe that the oscillation shape differs significantly from the $M < M_c$ regime (Fig.~\ref{fig:V}), especially for smaller force values. 
The impurity follows the metastable branch, overshoots it and then jumps off of it in a sharp fashion. For not too strong forces, the velocity takes multiple sharp turns, before it starts increasing parallel to the ground-state curve. The jump from the metastable branch becomes increasingly vertical in the limit of decreasing force, and occurs closer to its termination point.  
Periodic oscillations are established from the second oscillation for $\tilde{F}= 0.4$, and for higher magnitudes of force, such as $\tilde{F}=1$, from the third oscillation onward (see Fig.~\ref{fig:V8m}). Interestingly, there is an important deviation from the predicted dependence of the period on the force given by $T=2\pi\hbar n_0/F$. Already at $\tilde{F}=0.1$, the period is $18\%$ shorter than the expected one. Thus, the velocity displacement from the ground-state curve additionally increases after each oscillation (Fig.~\ref{fig:V8m}). 

We stress that the underlying mechanism of the oscillations is the same as for the case $M< M_c$, as explained in Sec.~\ref{sec:BlochMech}. The state realised along the jump from the metastable to the ground-state energy branch corresponds to the one occurring in the interval of forbidden momenta in Sec.~\ref{sec:BlochMech}, and is characterised by a depletion region attached to the density peak.

Figure \ref{fig:VDVB8m}a shows the behavior of the impurity drift velocity $V_d$ as a function of the external force. The drift velocity is an increasing function of the force in a large interval of forces. However, it exhibits a non-monotonicity in a narrow interval of forces near $\tilde{F}=0.6$. While the quantity $V_{\mathrm{min}}+V_{\mathrm{max}}$ monotonically increases with force, the shape of the oscillations is also changing with $F$, and causes this behavior of the drift velocity.
The slope and convexity of $V_d$ change several times in Fig.~\ref{fig:VDVB8m}. At small values of the force, $V_d$ grows rapidly and exhibits sub-linear dependence on $F$. In the intermediate range of forces it exhibits a considerably slower rate of change, while at large values of the force, $V_d$ again changes its slope and undergoes a faster, nearly linear, increase with $F$. For the force $\tilde{F}=3.5$ the Bloch oscillations do not take place anymore and the impurity undergoes an unlimited acceleration in time for this and higher values of the force.

Note that $V_d$ shows a tendency of a saturation into a non-zero value and the existence of a plateau at small magnitudes of the force. The latter can be understood from the energy dispersion relation.  Assuming that the impurity at $F\to 0^+$ follows its metastable branch till its termination point and then jumps to the ground-state branch while conserving the total momentum, the corresponding energy change $\Delta E$ is transferred to the host bosons. This process repeats over each oscillation and 
explains the saturation of the drift velocity curve to a nonzero value. However, for $\tilde{F}=0.03$ the value of the dissipated energy per period, $V_d T_{num} F$, is still around three times higher than $\Delta E$. Here $T_{num}$ is numerically obtained period of oscillations. Indeed, Fig.~\ref{fig:V8m} shows that the impurity velocity for $\tilde{F}=0.03$ differs from the above described scenario expected to occur for $F\to 0^+$, although showing a consistent trend toward it.

From \eq{eq:PolaronEnergy} follows that the energy difference $\Delta E$ increases with the impurity mass and saturates to $2\pi\hbar n_0 v_c$. Thus, the drift-velocity plateau at $F\to 0^+$ is expected to increase as the impurity become heavier, and to saturate to the critical velocity (\ref{eq:vc}) for $M\to \infty$ \cite{PhysRevLett.108.207001}.
Moreover, the quantum fluctuations allow for a tunnelling from the metastable branch to the ground state one. Their importance is controlled by the value of the inverse Luttinger liquid parameter $K^{-1}$. For weakly interacting bosons, $K^{-1}=\sqrt{\gamma}/\pi\ll 1$. Thus, the tunnelling plays an important role only for very small forces where the impurity momentum grows sufficiently slow in time such that the impurity spends sufficiently long time on the metastable branch that its probability to tunnel over this time becomes relevant. As a result, the saturation plateau in $V_d$ is expected to occur in the small-forces regime and to disappear in the limit of an asymptotically small force. This problem for a repulsive impurity-boson interaction has been studied in Ref.~\cite{PhysRevLett.108.207001}, leading to $V_d\sim F^{\alpha}$ with $\alpha\sim \sqrt{\gamma}$ for $F\to 0^+$. We expect that this scaling takes place also for $G<0$.

Note that for $M<M_c$ in the vicinity of the critical mass, the impurity dynamics exhibit similar drift velocity behavior and oscillation shapes within the regime of a small, but not asymptotically small, force. One example is the case of $G=-2$ and $M=2.5m$ considered in Fig.~\ref{fig:VdSmallF} in Sec.~\ref{sec:Bloch}.

The force dependence of the impurity velocity oscillation amplitude $V_B$ is shown in Fig.~\ref{fig:VDVB8m}b. It first increases, reaches a local maximal value and then decreases as a function of the force. Its slope changes considerably with force. Contrary to the results from the previous section, $V_B$ does not show a revival at high values of the force, but rather decays slowly with force.  At smallest considered value of the force, $V_B$ is $7.7\%$ smaller than the  above described prediction of small-force scenario. Thus, one would expect that $V_B$ changes again a trend and increases as $F$ decreases in the very small-$F$ regime.

Finally, we study the time period of oscillations in Fig.~\ref{fig:VDVB8m}c. Contrary to results for $M\leq M_c$, the period does not follow the usual $T=2\pi \hbar n_0/F$ relationship, and decreases in a faster manner with force. 
The only exception are very small forces. The relative error between the numerical and the expected time period for $\tilde{F}=0.035$ is $0.2\%$ of the expectation. For higher forces, the relative error grows, and already at $\tilde{F}=0.055$ and $\tilde{F}=1$ the period is $3\%$ and $35\%$ shorter than the expectation, respectively.

\section{Conclusions and discussion\label{sec:conclusions}}

We have studied the dynamics of an attractively interacting impurity driven by an external force through a bath of bosons under far-from-equilibrium conditions. We have shown that the impurity undergoes drifted Bloch oscillations in a wide interval of forces. The force threshold above which the impurity experiences an unlimited acceleration increases with the strength of the impurity-boson attraction.

Monitoring the time evolution of the boson density and the phase profile, we have characterized the dynamical response of the bosons. The impurity periodically transfers the momentum into the Bose gas in the form of density waves and solitons, as well as through the periodic increase of the phase variation along the system size. The key mechanism for the latter is the formation of a depletion region in the boson density attached to the density peak centred at the impurity position. Namely, in the interval of forbidden momenta, as the system momentum increases and the impurity velocity decays from its maximal value to zero, the depletion region forms and progressively deepens, reaching a complete depletion. The latter happens at $p=\pi \hbar n_0$ in the limit of very small external force. The impurity velocity changes its sign and the phase drop over the depleted region attached to it changes from $\pi$ into $-\pi$, leading to an increase of the phase variation over the system length by $2\pi$. This process repeats over each oscillation period.

Flipping the sign of the impurity-boson interaction strength $G$ from repulsive to attractive significantly alters the system dynamics.
For $G>0$, the soliton-like stationary states (\ref{densityStationary}) are realised  for any system momentum in the limit of an infinitesimal force, and the drift velocity depends linearly \cite{AnnalsKamenev} on the force in a wide interval of forces \cite{BlochAP}. 
For $G<0$, rather than a linear dependence, we find a sub-linear dependence of the impurity drift velocity on the force in the weak-force regime relevant for experimental realisations. Namely, $V_d/v\sim \tilde{F}^{\alpha}$ with a positive exponent $\alpha$ being smaller than unity. Moreover, the exponent $\alpha$ appears to be non-universal and shows a tendency to decrease with the absolute value of the impurity coupling constant. Note that to characterize the small-$F$ regime for $G<0$, we studied force values an order of magnitude smaller than those in Ref.~\cite{BlochAP}, where we investigated the case $G>0$. 

Increasing the driving force, more energetic excitations are created, leading to a monotonic increase of the drift velocity. However, its functional form, convexity, and rate of this increase vary considerably with $F$. 
The dependence on the strength of the impurity-boson attraction is also rich. At low $F$ values, the drift velocity tends to increase with the absolute value of $G$; however, this trend reverses as $F$ increases.
The Bloch amplitude $V_B$ of velocity oscillations is a non-monotonic function of the force: first, $V_B$ decreases and reaches a local minimum, and then it grows with increasing force. 

The underlying mechanism for the Bloch oscillations described above also takes place for $M>M_c$. However, while the oscillation period matches $T = 2\pi \hbar n_0/F$ over a wide range of forces for $M\leq M_c$, this prediction is valid only at very weak forces for $M>M_c$.
For an impurity that exceeds the critical mass, the velocity oscillations significantly change in shape, particularly at lower force values. The impurity tracks the metastable branch, overshoots it, and then drops off sharply. As the force decreases, this drop becomes increasingly vertical and occurs closer to the termination point of the branch. 
Thus, in the small-force regime, the impurity saturates into a nonzero drift velocity as force decreases. However, at extremely small forces, in the limit $F\to 0^+$, quantum tunnelling becomes relevant and modifies this behavior into $V_d\sim F^{\alpha}$, where $\alpha\sim \sqrt{\gamma}$ \cite{PhysRevLett.108.207001}.  

Cold atomic gases provide an ideal testbad for the impurity dynamics studied in this work \cite{2012quantum,Meinert945,JinPhysRevLett.117.055301,PhysRevLett.117.055302,HadzibabicPhysRevX.15.021070,grusdt2025impurities,Polaron2D}.

\section{Acknowledgments}

This study has been partially supported through the EUR grant NanoX n° ANR-17-EURE-0009 in the framework of the ``Programme des Investissements d'Avenir".


%

\end{document}